\documentclass[superscriptaddress,secnumarabic,
amssymb,amsmath,nobibnotes,aps,prd,showkeys,showpacs,nofootinbib]{revtex4}%
\usepackage{graphicx}
\usepackage{epsf}
\usepackage{bm}
\usepackage{amsmath}
\usepackage{amsfonts}
\usepackage{amssymb}
\usepackage{epstopdf}
\usepackage{natbib}%
\providecommand{\U}[1]{\protect\rule{.1in}{.1in}}
\newcommand{\be}{\begin{equation}}
\newcommand{\ee}{\end{equation}}

\newcommand{\mincir}{\raise
-3.truept\hbox{\rlap{\hbox{$\sim$}}\raise4.truept\hbox{$<$}\ }}
\newcommand{\magcir}{\raise
-3.truept\hbox{\rlap{\hbox{$\sim$}}\raise4.truept\hbox{$>$}\ }}

\usepackage{color}

\begin{document}
\title{Gravitationally induced particle production and its impact on structure formation}
\author{Rafael C. Nunes}
\email{rafadcnunes@gmail.com}
\affiliation{Departamento de F\'isica, Universidade Federal de Juiz de Fora, 36036-330, Juiz de Fora, MG, Brazil}


\begin{abstract}

\noindent  In this paper we investigate the influence of a continuous particles creation processes on  
           the linear and nonlinear matter clustering, and its consequences
           on the weak lensing effect induced by structure formation. 
           We study the line of sight behavior of the contribution 
           to the bispectrum signal at a given angular multipole $l$, showing that the scale 
           where the nonlinear growth overcomes the linear effect depends strongly 
           of particles creation rate.
           
\end{abstract}

\maketitle

\section{Introduction}
\label{sec:intro}

The matter creation in an expanding universe is not a new issue in cosmology. 
Such physical mechanism has been intensively investigated. Zeldovich \cite{Zeldovich} describes 
the process of creation of matter in the cosmological context through of an effective mechanism. 
Irreversible processes also were investigated in the context of inflationary scenarios \cite{Turner, Barrow, Turok, Starobinsky,
ccdm8,ccdm7,ccdm10,ccdm13,ccdm14}.
Prigogine \cite{Prigogine} studied how to insert the creation of matter consistently in Einstein's field equations.
Many authors have explored scenarios of matter creation in cosmology, but here we are particularly interested in
the gravitationally induced particle creation scenario denominated, creation of cold dark matter (CCDM) \cite{ccdm1,ccdm2,
ccdm3,ccdm4,ccdm5,ccdm6,ccdm9,ccdm11,ccdm12,ccdm15,ccdm16}.
\\

Recent determinations of the equation of state of dark energy hint that this may well be of the phantom
type, i.e., $w_{de} < -1$ \cite{ade,Rest,Xia,Cheng,Shafer,Conley,Scolnic}. If confirmed by future experiments, 
this would strongly point to the existence of fields that violate the dominant energy condition, 
which are known to present serious theoretical difficulties \cite{Caldwell,Carroll,Cline,Hsu,Sbisa,Dabrowski}.
In a recent work \cite{Dr1} we have investigated an alternative to this possibility, namely, that
the measured equation of state, $w_{de}$ , is in reality an effective one, the equation of state
of the quantum vacuum, $w_{\Lambda} = -1$, plus the negative equation of state, $w_c$, associated to
the production of particles by the gravitational field acting on the vacuum. Thus, 
in this new scenario, the effective equation of state comes to be $w_{eff}= w_{\Lambda} + w_c < -1$. 
In this paper, we aim is to explore the consequences of matter creation processes 
on the linear and nonlinear perturbations of matter, considering the combined effect of $w_{\Lambda} + w_c$ 
on the process of structure formation.
\\

This paper is organized as follows. Section II we will make a brief review of the 
cross-power spectrum of the lensing potential-Rees-Sciama effect. Section III 
briefly sums up the phenomenological basis of particle creation in expanding 
homogeneous and isotropic, spatially flat, universe. Section IV,
we evaluate the consequences of the matter creation process on linear 
and nonlinear power spectrum. Section V, we discussed the lensing-Rees-Sciama power spectrum. 
Lastly, Sec. VI briefly delivers our main conclusions and offers some final 
remarks. As usual, a zero subscript means the present value of the corresponding quantity.


\section{Weak lensing-Rees-Sciama}

The path of the CMB photons traveling from the last scattering surface can be modified by the gravitational 
fluctuations along the line-of-sight. On angular scales much larger than 
arcminute scale the photon's geodesic is deflected by gravitational
lensing and late time decay of the gravitational potential and nonlinear growth 
induce secondary anisotropies known respectively as the Integrated Sachs Wolfe (ISW) \cite{ISW} and 
the Rees-Sciama (RS) effect \cite{RS}. Hereafter by RS we refer to the combined contribution of 
linear and nonlinear growth.
\\ \

The CMB anisotropy in a direction \textbf{\^{n}} can then be decomposed into:

\begin{equation}
\label{CMB_total}
\Theta (\textbf{\^{n}}) = \Theta^P (\textbf{\^{n}})+\Theta^L (\textbf{\^{n}})+\Theta^{RS} (\textbf{\^{n}}), 
\end{equation} 
where $P$ denotes primary, $L$ lensing and $RS$ ISW+Rees-Sciama, which includes both the linear and the nonlinear contributions. 
This last term takes the form

\begin{equation}
\label{efeito_RS}
\Theta_{RS} (\textbf{\^{n}})= 2 \int dr \frac{ \partial }{ \partial \eta } \Phi( \eta, \textbf{\^{n}} r),
\end{equation}
where $\eta$ is the cosmological conformal time, $r$ the comoving distance along the line of sight, 
$r(z)=1/H_0 \int_0^z dz'/E(z)$, and $\Phi$ is the cosmological gravitational potential, defined as the 
fluctuation in the metric  \cite{ISW,Hu}.
\\

Note that in general Eq. (\ref{efeito_RS}) describes the contribution from linear and nonlinear density 
fluctuations, with the only assumption that the gravitational potential fluctuations they induce are linear. 
The effect of gravitational lensing on the CMB is contained in the second term of Eq. (\ref{CMB_total}),
where the weak lensing on the CMB  is qualified by, $\Theta^L (\textbf{\^{n}})= \Theta^P (\textbf{\^{n}}+ \nabla \phi) 
\simeq \Theta^P (\textbf{\^{n}}) + \nabla_{i} \phi(\textbf{\^{n}})$ \cite{Verde}.
\\

The deflection angle $\alpha=\nabla \phi_L$ is given by the angular gradient of the gravitational 
potential projection along the line of sight

\begin{equation}
\label{}
\phi_L (\textbf{\^{n}})= -2 \int_0^{z_{ls}} dr \frac{r_{ls}-r}{r_{ls} \, r} \Phi(r,\textbf{\^{n}}r),
\end{equation}
where $r_{ls} = r(z_{ls})$ refers to the comoving distance to the last scattering from the observer at $z = 0$.
\\

Note that the RS and lensing effects are correlated since they both arise from $\Phi$; this correlation induces a 
non-Gaussian feature on the CMB pattern and leads to a non vanishing bispectrum signal \cite{Hu,Verde,Komatsu}.
The CMB bispectrum is built out of the third order statistics in the harmonic domain, 
and for lensing-RS correlation is given by \cite{Komatsu,Goldberg}

\begin{eqnarray}
\label{}
B_{ l_1 l_2 l_3}^{ m_1 m_2 m_3}= <a_{l_1 m_1} a_{l_2 m_2} a_{l_3 m_3} > = G_{ l_1 l_2 l_3}^{ m_1 m_2 m_3} \Big[ \frac{l_1(l_1+1)-l_2(l_2+1)+l_3(l_3+1)}{2} \times  \nonumber \\ 
C_{l_1} <\Theta^{*}_{l_3 m_3} a^{RS}_{l_3 m_3}> + 5 \, \, perm \Big],
\end{eqnarray}
where for Gaussian fields, expectation value of the bispectrum is exactly zero.
\\ 

Here, $C_l$ is the primary CMB power spectrum with lensing, and $a^{RS}_{lm}$ are the spherical 
harmonics coefficients of the RS effect. The Gaunt integral, $G_{ l_1 l_2 l_3}^{ m_1 m_2 m_3}$, is defined as 

\begin{equation}
\label{}
G_{ l_1 l_2 l_3}^{ m_1 m_2 m_3}= \sqrt{\frac{(2l_1+1)(2l_2+1)(2l_3+1)}{4 \pi}} \left(\begin{array}{rrr} l_1&l_2&l_3\\ 0&0&0 \end{array}\right) \left(\begin{array}{rrr} l_1&l_2&l_3\\ m_1&m_2&m_3 \end{array}\right).
\end{equation}
\\

Assuming statistical isotropy of the universe, rotational invariance implies that one can average over orientation 
of triangles (i.e., $m's$) to obtain the angle-averaged bispectrum \cite{Spergel}

\begin{eqnarray}
 \label{}
 B_{ l_1 l_2 l_3}= \sum_{m_1 m_2 m_3}  \left(\begin{array}{rrr} l_1&l_2&l_3\\ m_1&m_2&m_3 \end{array}\right) B_{ l_1 l_2 l_3}^{ m_1 m_2 m_3}=  
 \sqrt{\frac{(2l_1+1)(2l_2+1)(2l_3+1)}{4 \pi}} \left(\begin{array}{rrr} l_1&l_2&l_3\\ 0&0&0 \end{array}\right) \times  \nonumber \\ 
 \Big[ \frac{l_1(l_1+1)-l_2(l_2+1)+l_3(l_3+1)}{2} 
 C_{l_1} <\Theta^{*}_{l_3 m_3} a^{RS}_{l_3 m_3}> + 5 perm \Big]
\end{eqnarray}


Here, the cross-power spectrum of the lensing potential RS effect, $Q(l)=<\Theta^{*}_{l_3 m_3} a^{RS}_{l_3 m_3}>$, 
is given by \cite{Goldberg,Verde}

\begin{equation}
\label{Ql}
Q(l)=<\Theta^{*}_{l_3 m_3} a^{RS}_{l_3 m_3}>=2 \int_0^z dz \frac{r(z_{ls})-r(z)}{r(z_{ls})r(z)^3} \frac{\partial P_{\Phi} (k,z)}{\partial z} |_{k=l/r(z)},
\end{equation}
where $P_{\Phi}$ is the power spectrum of the Newtonian potential

\begin{equation}
\label{}
P_{\Phi}(k,z)=\frac{9}{4} \Omega^2_{m} \Big(\frac{H_0}{k} \Big)^4 (1+z)^2 P(k,z),
\end{equation}
and $P(k,z)$ is the power spectrum of matter density fluctuations. 
\\ \

The quantity more relevant here is, $Q(l)$, describing how the forming structures along the 
line of sight induce the lensing on CMB photons, expressed as the statistical expectation of the 
correlation between the RS and lensing effects. In particular, the expression above has been used to 
evaluate the bispectrum dependence on the most important cosmological parameters, including an effectively 
constant dark energy equation of state, and the benefits of the bispectrum data on the estimation of 
the cosmological parameters themselves \cite{Verde,Fabio}.

\section{Cosmological models with particle creation}\label{sec:models}

\noindent As investigated  by Parker and collaborators
\cite{Parker}, the material content of the Universe may have had
its origin in the continuous creation of radiation and matter from
the gravitational field of the expanding cosmos acting on the
quantum vacuum, regardless of the relativistic theory of
gravity assumed. In this picture, the produced particles draw
their mass, momentum and energy from the time-evolving
gravitational background which acts as a ``pump" converting
curvature into particles.
\\

\noindent Prigogine  \cite{Prigogine} studied how to insert the
creation of matter consistently in Einstein's field equations.
This was achieved by introducing in the usual balance equation for
the number density of particles, $ (n\, u^{\alpha})_{; \alpha}=0$,
a source term on the right hand side to account for production of
particles, namely,
\begin{equation}
(n\, u^{\alpha})_{; \alpha} = n \Gamma \, ,
\label{Eq:nbalance}
\end{equation}
where $u^{\alpha}$ is the matter fluid four-velocity normalized so
that $ u^{\alpha}\, u_{\alpha} = 1$ and $\Gamma$ denotes the
particle production rate. The latter quantity essentially vanishes
in the radiation dominated era since, according to Parker's theorem, the production of
particles is strongly suppressed in that era \cite{Parker-Toms}.
The above equation, when combined with the second law of
thermodynamics naturally leads to the appearance of a negative
pressure directly associated to the rate  $\Gamma$, the creation
pressure $p_{c}$, which adds to the other pressures (i.e., of
radiation, baryons, dark matter, and vacuum pressure) in the total
stress-energy tensor. These results were subsequently discussed
and generalized in \cite{Lima1}, \cite{mnras-winfried}, and
\cite{Zimdahl} by means of a covariant formalism, and  further
confirmed using relativistic kinetic theory
\cite{cqg-triginer,Lima-baranov}.
\\  \

\noindent Since the entropy flux vector of matter, $n \sigma
u^{\alpha}$, where $\sigma$ denotes the entropy per particle, must
fulfill the second law of thermodynamics $(n \sigma u^{\alpha})_{;
\alpha} \geq 0$, the constraint $\Gamma \geq 0$ readily follows.
\\  \

\noindent For a homogeneous and isotropic universe, with scale
factor $a$, in which there is an adiabatic process of particle
production from the quantum vacuum  it is easily found that
\cite{Lima1,Zimdahl}
\begin{equation}
\label{pressure_creation} p_c= - \frac{\rho \, + \, p}{3H}\,
\Gamma \, .
\end{equation}
Therefore, being $p_{c}$ negative it can help drive the era of
accelerated cosmic expansion we are witnessing today. Here $\rho$
and $p$ denote the energy density and pressure, respectively,  of
the corresponding fluid, $H=\dot{a}/a$  is the Hubble factor, and,
as usual, an overdot denotes differentiation with respect to
cosmic time. Since the production of ordinary particles is much
limited by the tight constraints imposed by local gravity
measurements \cite{plb_ellis, peebles2003, hagiwara2002}, and
radiation has a negligible impact  on the recent cosmic dynamics,
for the sake of  simplicity, we will assume  that the produced
particles are just dark matter particles.
\\ \

\noindent Let us consider a spatially flat FRW universe dominated
by pressureless matter (baryonic plus dark matter) and the energy
of the quantum vacuum (the latter with EoS $p_{\Lambda} = -
\rho_{\Lambda}$) in which a process of dark matter (DM) creation from
the gravitational field, governed  by
\begin{equation}
\label{dark_matter_evolution} \dot{\rho}_{dm} \,  + \, 3H
\rho_{dm} = \rho_{dm} \, \Gamma
\end{equation}
is taking place. In writing the last equation, we used directly the Eq. (\ref{Eq:nbalance}) specialized to dark matter particles 
and the fact $\rho_{dm} = n_{dm} \, m$, where $m$ stands for the rest mass
of a typical dark matter particle \cite{Prigogine,Lima1,mnras-winfried}.
Since baryons are neither created nor destroyed, their corresponding energy density obeys
$\dot{\rho}_{b} \, + \, 3H \rho_{b} = 0$. On their part, the
energy of the vacuum does not vary with expansion, hence $\,
\rho_{\Lambda} = {\rm constant}$. Thus, Friedmann equation for this scenario
can be expressed simply as 

\begin{equation}
\frac{H^2(a)}{H^2_0}=\Omega_{b0}\, a^{-3} \, +\, \Omega_{dm0}  a^{-3} \Big( \exp{\int_1^a \frac{da}{a}\frac{\Gamma}{H}} \Big) +
 \, \Omega_{\Lambda0}.
 \label{Eq:Hz}
\end{equation}
\\

To go ahead an expression for the rate $\Gamma$ is needed.
However, the latter cannot be ascertained before the nature of
dark matter particles be discovered. Thus, in the meantime, we
must content ourselves with phenomenological expressions of
$\Gamma$. In \cite{Dr1} was proposed three parameterizations 
for investigate the dynamic consequences of the matter creation, 

\begin{equation}
\label{proposta1} \qquad \qquad \Gamma = 3 \beta H  \qquad \qquad
\qquad \; \; \; {\rm (Model \, I)},
\end{equation}
\begin{equation}
\label{proposta2} \Gamma = 3 \beta \, H \,  [5-5 \tanh (10-12a)]
\qquad {\rm (Model \, II)},
\end{equation}
and
\begin{equation}
\label{proposta3} \Gamma = 3 \beta \, H \, [5-5 \tanh (12a-10)]
\qquad {\rm (Model \, III) }.
\end{equation}
\\

\begin{figure}
   \includegraphics[width=4in, height=3in]{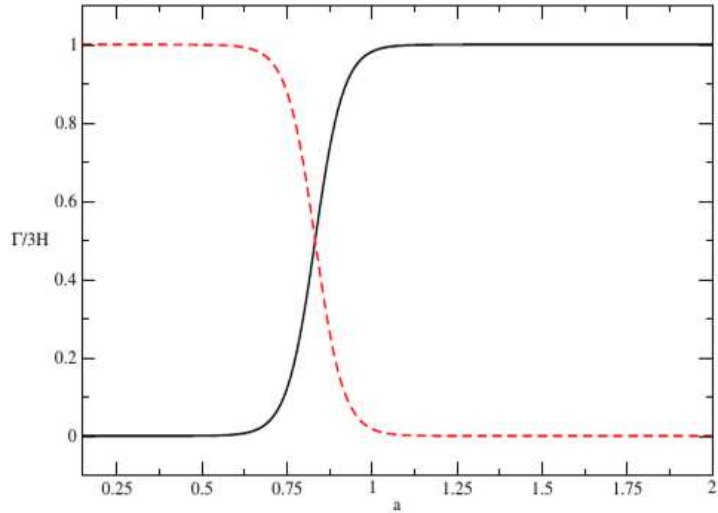}
   \caption{\label{Gamma_3H} Evolution of ratio $\Gamma/3H$ as a function of the scale factor for the model II (solid, black
line) and model III (dashed, red line). In drawing the graphs we have taken $\beta$ = 0.1. The ratio $\Gamma/3H$ for the models I, 
is a constant equal to 0.1.}
\end{figure}

Figure \ref{Gamma_3H} shows the said ratio $\Gamma/3H$ in terms of the scale
factor for the models II and III assuming $\beta = 0.1$. It should be noted that for $\beta$ larger than 0.1 one is led to $w_{eff} < -2$, 
which lies much away from the reported w de values. In all three cases, $\Gamma/3H \leq 1$ at any scale factor for 
$\beta \leq 0.1$. In \cite{ccdm9} is demonstrated through generalized second law of thermodynamics 
bounded by the apparent horizon that the particle production rate is limited by $\beta \leq 0.1$ for the model I, and $\beta \leq 1$ 
for the model II and III.
The main characteristic of this new scenario proposed in \cite{Dr1}, is that the measured equation of state (EoS) 
is in reality an effective one, the EoS of the quantum vacuum, $w_{\Lambda} = -1$, plus the negative EoS, $w_c$, 
associated to the production of particles by the gravitational field acting on the vacuum.
Thus, the effective EoS comes to be $w_{eff} = w_{\Lambda} + w_c < - 1$, where
$w_c$ is the EoS related to the creation pressure Eq. (\ref{pressure_creation}). 
If the answer is affirmative, then the need to recourse to phantom
fields will weaken \cite{Caldwell, Carroll,Cline, Hsu, Sbisa, Dabrowski}. 
Therefore, the real motivation these models is explain a phantom behavior without 
the need of invoking scalar fields and any modified gravity.
From a joint analysis of Supernova type Ia, gamma ray bursts, baryon acoustic oscillations, and the Hubble rate, 
it was obtained $w_{eff}(z=0) = -1.073^{+0.034}_{-0.035}$, $-1.155^{+0.076}_{-0.080}$, and $-1.002^{+0.001}_{-0.001}$ 
for models I, II, and III, respectively at 1$\sigma$ $-$ see \cite{Dr1} for 
more details on the motivation of the models Eqs (\ref{proposta1})-(\ref{proposta3}), 
and information of the statistical results obtained for the $\beta$ parameter. 
The next section, let us  investigate the consequences of this model on the process of structure formation  
in small and large scales. 

\section{Growth of Perturbations}
\label{sec:GP}

\subsection{Linear effects}

\noindent In the linear theory - valid at sufficiently early times and large spatial scales, when the
fluctuations in the matter-energy density are small- the density
contrast of matter, $\delta = \delta \rho_m / \rho_m$, evolves
independently of the spatial scale of the perturbations. R. Reis
\cite{reis}, and more recently, O. Ramos {\it et al.} \cite{ramos} 
describes the evolution of linear matter perturbations in the case of models of continuous
matter creation, with rate $\Gamma$. The growth of fluctuations follows by integrating equation (4.8) in \cite{ramos},
written under the understanding of dealing with pressureless matter 

\begin{equation}
\label{delta_wc}
 \delta_{m}''+ \frac{3}{2a}(1-5w_c)\delta_{m}'+\frac{3}{2a^2}(3w_c^2-8w_c-1)\delta_{m}=0,
\end{equation}
where the prime denotes derivative with respect to scale factor $a$, and $w_c$ represents 
the EoS parameter associated with the matter creation processes.
Note that for $w_c=0$ , there is not matter creation and the model reduces to the Einstein-deSitter.
The introduction of the contribution of dark energy for the growth of linear perturbations 
of matter can be taken with developed by Linder \cite{linder1} (see also \cite{linder2,wmap5} for more details).
Thus, following the methodology developed in \cite{linder1}, the Eq. (\ref{delta_wc}) can be rewritten as 

\begin{equation}
\label{delta_wc2}
 \delta_{m}''+ \frac{3}{2a} \Big(1-5w_c-\frac{w}{1+X(a)} \Big)\delta_{m}'+\frac{3}{2a^2}\Big(3w_c^2-8w_c-\frac{X(a)}{1+X(a)} \Big)\delta_{m}=0,
\end{equation}
where

\begin{equation}
 X(a)= \frac{\Omega_m}{1-\Omega_m} e^{-3 \int d \ln a' w},
\end{equation}
and for the cosmological scenario presented in this work, $w=w_{\Lambda}=-1$. Note that for $w_c=0$, we obtain the standard equation that 
describes the evolution of the linear perturbations of matter for the $\Lambda$CDM model, and with
$w=w_c=0$, the Eq. (\ref{delta_wc2}) reduces to the Einstein-deSitter. 
\\

\begin{figure}
   \includegraphics[width=5in, height=3in]{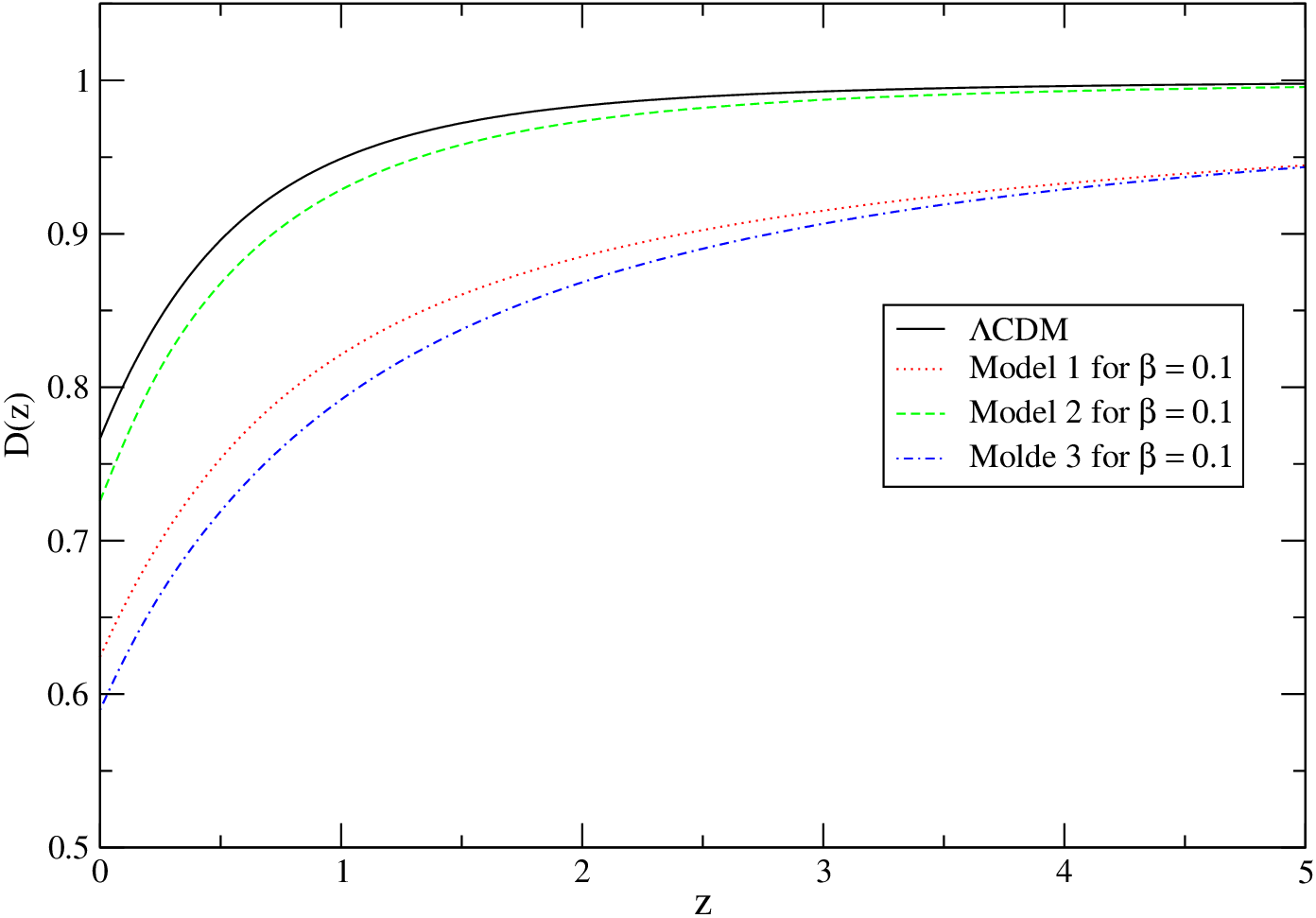}
   \caption{\label{delta_model1} The matter density constrast, $D(z)=(1+z)\delta(z)/\delta(z=0)$, for the models I, II, III, and $\Lambda$CDM,  
   as a function of the redshift $z$ for $\beta = 0.1$.}
\end{figure}

Figure \ref{delta_model1} shows the evolution of linear perturbations of matter in terms of $D(a)= (1 +z)\delta(z)/\delta(z=0)$ 
for the models I, II, and III assuming $\beta=0.1$. Note that the density contrast is suppressed as we increase the 
rate production of matter.
\\

The linear growth function $\delta(a)$ can be parameterized of efficiently through the introduction of growth index

\begin{equation}
f \equiv \frac{d \ln \delta(a)}{d \ln a}=\Omega^{\gamma}_m(a),
\end{equation}
where $a=1/1+z$.

This parameterization was originally introduced by Peebles \cite{Peebles} and then by Wang and Steinhardt \cite{growth_index}, 
and it was shown to provide an excellent fit corresponding to various general relativistic 
cosmological models for specific values of $\gamma$. We will use 11 data points reported in \cite{growth_data}, 
to estimate the model parameters by minimizing the quantity

\begin{equation}
\label{chi_quadrado_f}
\chi^2_{f}(\theta_{i})=\sum_{i=1}^{11}\frac{[f^{obs}(z_i)-f^{th}(z_i,\theta_{i})]^2}{\sigma^2(z_i)},
\end{equation}
where $\theta_{i}= \{\beta, \gamma \}$  and the set of free parameters, and $\sigma(z_i)$ the error observed in 1$\sigma$.

\begin{figure}
   \includegraphics[width=5in, height=3in]{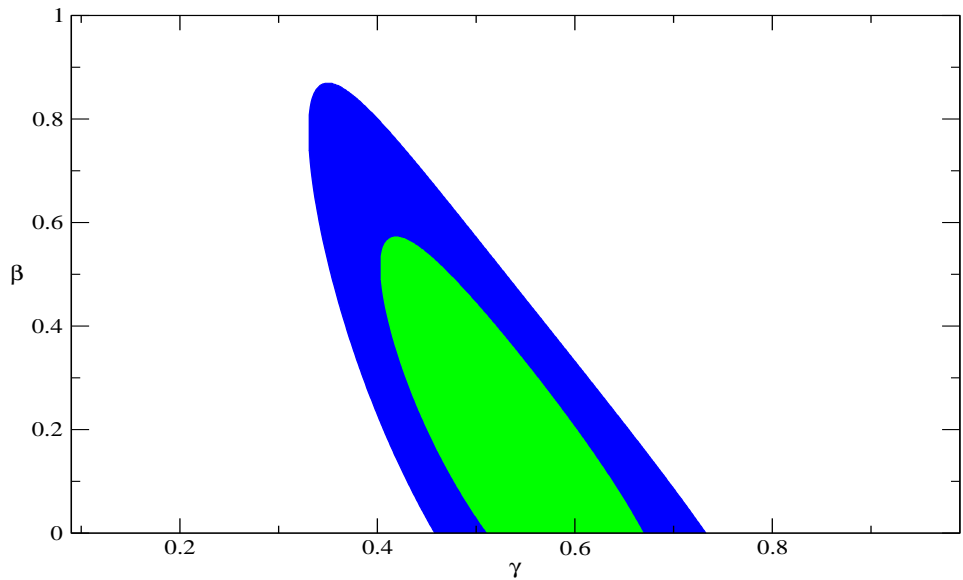}
   \caption{\label{growth1} 1$\sigma$ and 2$\sigma$ confidence regions in the plan $\gamma-\beta$ for the model 1.}
\end{figure}

\begin{figure}
   \includegraphics[width=5in, height=3in]{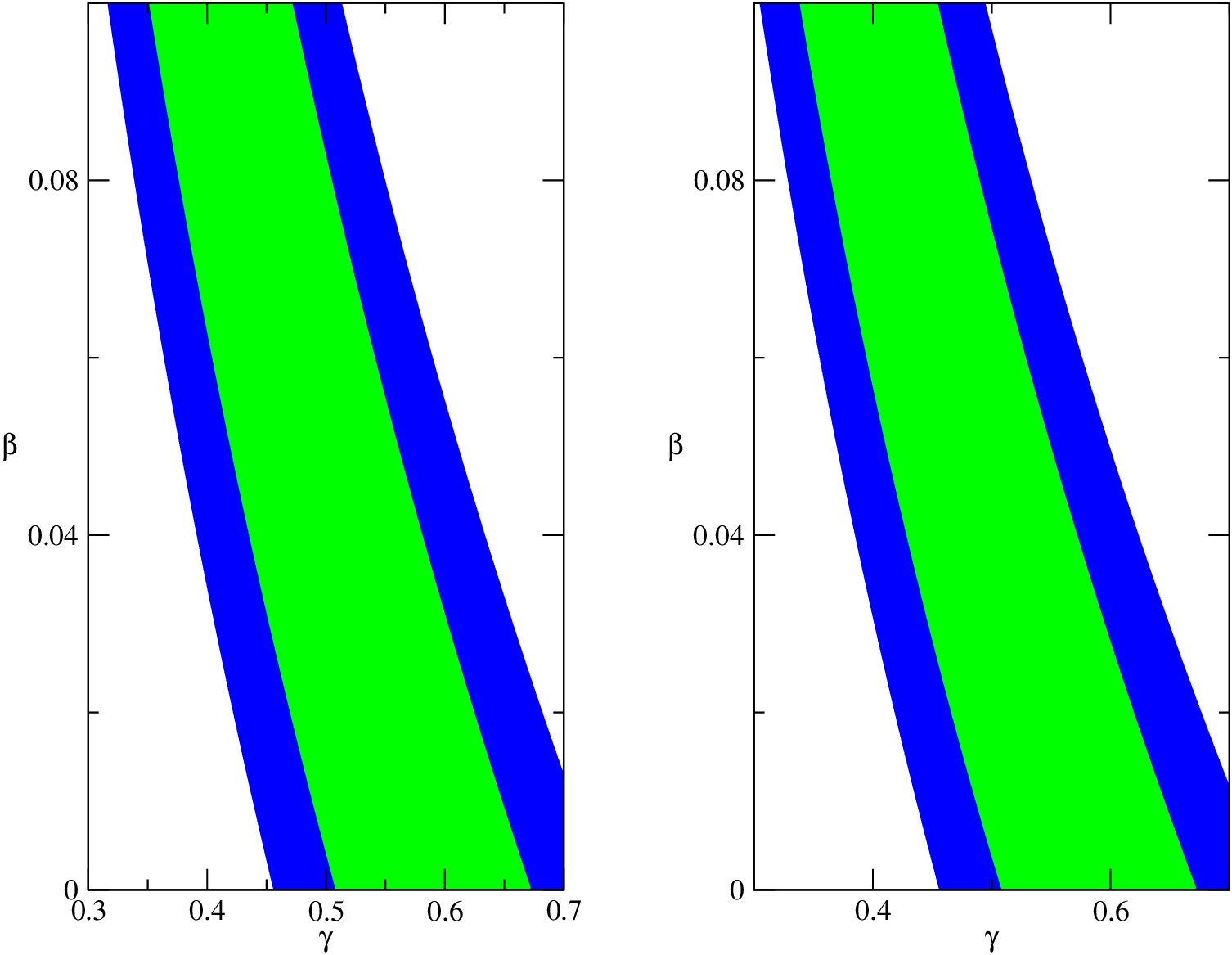}
   \caption{\label{growth23} Thus as in the figure \ref{growth1}, but for the model II (left panel)
  and model III (right panel).}
\end{figure}

Figure \ref{growth1} and \ref{growth23} shows the confidence regions at 1$\sigma$ and 2$\sigma$ for model I, and models II and III,
respectively. We note with best fits for Model I, $\gamma = 0.536^{+0.162}_{-0.132}$  and $\beta = 0.160^{+0.412}_{-0.160}$. 
For models II (III) we note that $\gamma = 0.4093^{+0.262}_{-0.057} (0.3944^{+0.277}_{-0.055})$ and $\beta = 0.10^{+0.0}_{-0.10} (0.100^{+0.0}_{-0.1})$. The upper and lower values 
represent the errors at 1$\sigma$. During the statistical analysis we assumed $\beta \in [0,1]$ for model I, 
and $\beta \in [0,0.1]$ for models II and III. These are the boundaries for the parameter $\beta$ imposed by thermodynamics \cite{ccdm9}.
\\

The real space galaxy power spectrum is related to the matter power spectrum as follows, $P_g=b^2(z)P_m$, 
where $b(z)$ is the bias factor between galaxy and matter distributions. The matter power spectrum is defined as

\begin{equation}
P_m(k)=\mid\delta_k\mid^2,
\end{equation}
where $\delta_k$ is the Fourier transform of the matter density perturbation, defined as

\begin{equation}
\delta_k \equiv \int \delta(r) e^{ik \cdot r} dr.
\end{equation}

\noindent  To calculate the power spectrum of galaxies of the models presented in this paper, we modified the program 
developed by E. Komatsu \cite{komatsu} that calculate the spherically-averaged power spectrum. Results on the 
clustering of 282.068 galaxies in the Baryon Oscillation Spectroscopic Survey (BOSS) show that  at large scales, 
the bias factor and measured in $b$ = 2.00 $\pm$ 0.07 \cite{SDSS}. We use this value to calculate the power spectrum 
for the models presented.
\\ 

\begin{figure}
  \includegraphics[width=5in, height=3in]{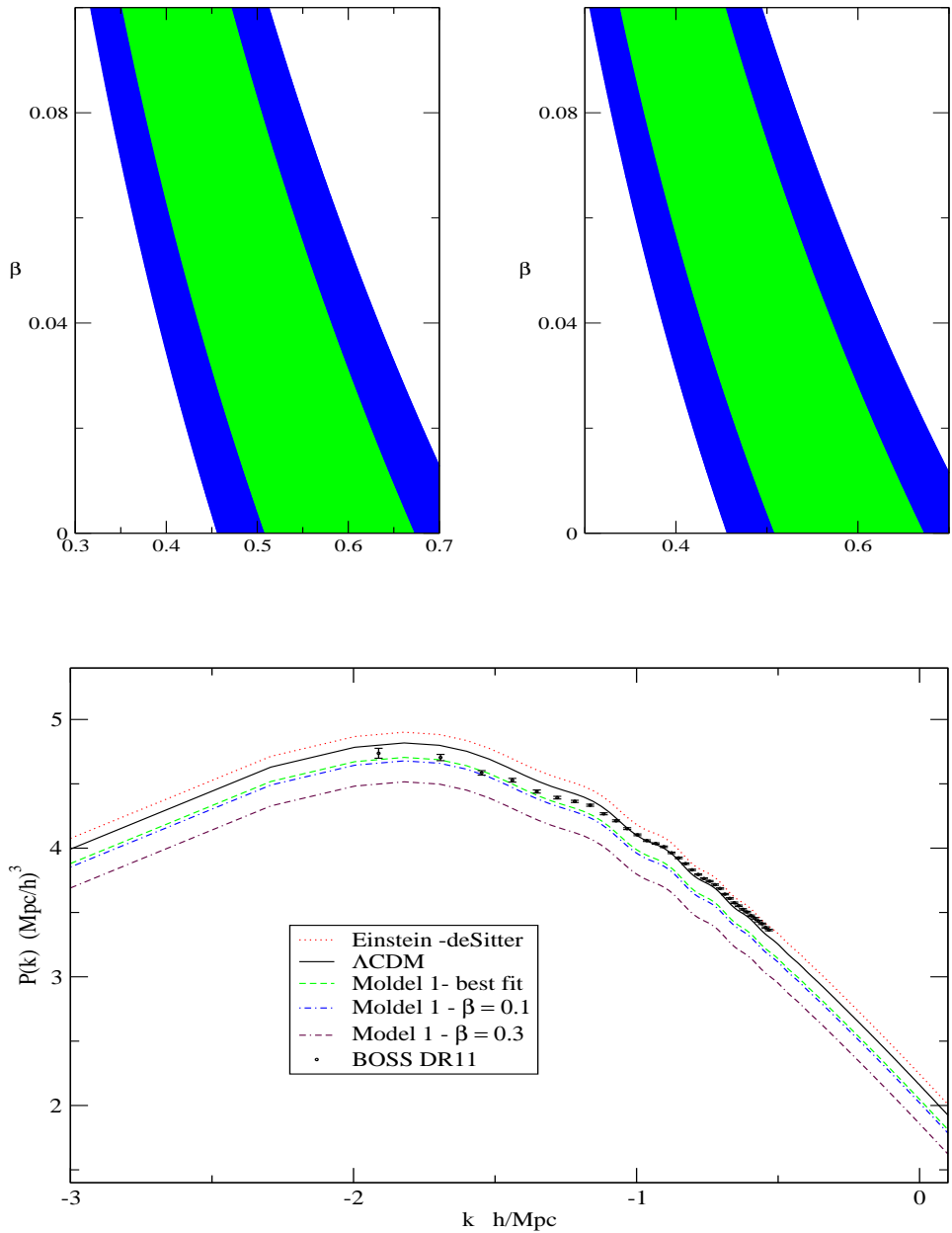}
  \caption{\label{pk_model1} Power spectrum for the model I for different values of $\beta$. Also we shows the
  theoretical prediction for the Einstein- de Sitter and $\Lambda$CDM models.
  In drawing the graphs we have assumed $\Omega_{m0} = 0.28$.
  The data with their erro bars were taken from \cite{BOSS}.}
\end{figure}

\begin{figure}
  \includegraphics[width=6.5in, height=3.5in]{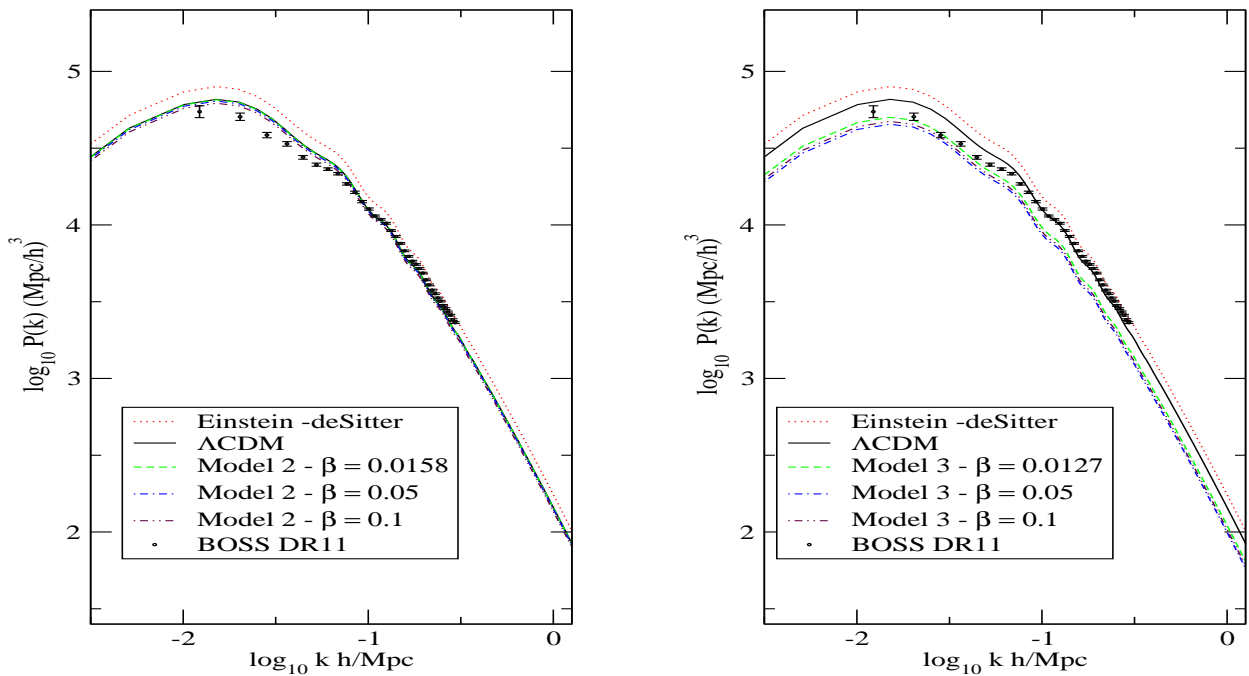}
  \caption{\label{pk_model23} Thus as in the figure \ref{pk_model1}, but for the model II (left panel)
  and model III (right panel).}
\end{figure}

\noindent  We compared the theoretical prediction for the power spectrum of galaxy with the data of 
BOSS DR11 \cite{BOSS}. Figure \ref{pk_model1} show the result for model I. Again, we see that the density contrast 
suppression as we increase the rate production of matter (high values of $\beta$), 
when compared to the $\Lambda$CDM. Figure \ref{pk_model23} shows the power spectrum 
for the models II (left panel) and for the model III (right panel). 
\\

\subsection{Non-linear effects}

The growth of inhomogeneities on small scale are highly non-linear, with density contrast of approximately
$\delta \sim 100$ for $z \ll 1$.  Galaxies form at redshifts of the order of 2-6; clusters of galaxies form in
redshifts the order of 1, and superclusters are forming just now, then the nonlinear effects are important for
understanding the universe at the level of clusters and superclusters of galaxies.
\\

On larger scales than 100 Mpc, where collapse is yet to come, the cosmological theory of linear disturbances
can be used directly. On small scales, the only safe way to perform calculations beyond the capacity of the
linear theory and through numerical simulations. Analytical approaches can offer insights very useful for
investigating the nonlinear stage of the inhomogeneities. Jeong and Komatsu \cite{non_linear_Pk1},
proposed an approximation in third order of perturbation (3PT) for the nonlinear matter power spectrum as following

\begin{equation}
\label{Pk_nonlinear}
P(k,z)=D^2(z)P_{11}(k)+D^4(z)[ P_{22}(k)+2P_{13}(k)],
\end{equation}
where $D(z)$ is a suitably normalized linear growth factor (which is proportional to the scale factor during
the matter era), $P_{11}(k)$ is the linear power spectrum at an arbitrary initial time, $z_i$,
at which $D(z_i)$ is normalized to unity, and $P_{22}(k)$ and $P_{13}(k)$ are both correction to the density (matter) power spectrum. \cite{non_linear_Pk1,non_linear_Pk2}. 
This is an attractive approach, as it provides the exact calculation in the quasi linear regime where the
perturbative expansion is still valid. Other approaches as Halo Model \cite{halo_model} and HKLM Scaling Model \cite{Hamilton} are
calibrated using numerical simulations with a specific set of cosmological parameters, and thus cannot be
easily extended to other cosmological models, such as dynamical dark energy models. A disadvantage of the
3PT is that its validity is limited to the quasi linear regime, and thus the result on very small scales cannot
be trusted. Thus, under this approach we apply this method to investigate the effects of nonlinear matter
clustering for models with creation of matter. 

\begin{figure}
\includegraphics[width=6in, height=3in]{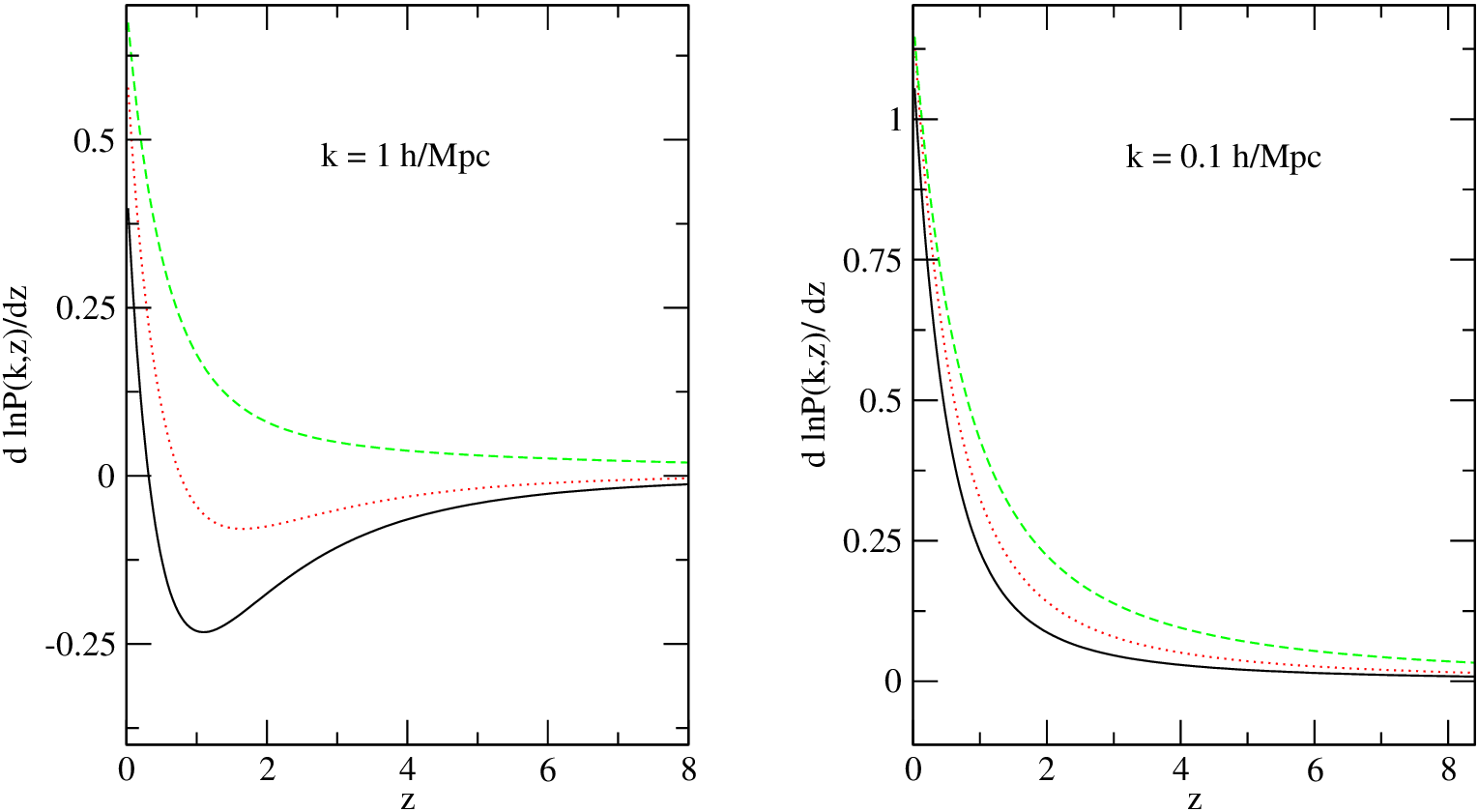}
\caption{\label{dPdz_model1} Left panel: Evolution of nonlinear power spectrum for the model I at $k = 1$ h$Mpc^{-1}$, 
 where the solid (black), dot (red), and dashed (green) lines represent $\beta = $ 0.025, 0.05, and 0.1, respectively. 
 Right panel : Same that in the left panel, but at $k = 0.1$ h$Mpc^{-1}$.}
\end{figure}

Let us now calculate the evolution of nonlinear power spectrum as a function of redshift, $\partial P_{\Phi} (k,z)/\partial z$.
This is related to derivatives of the density power spectrum, $P(k,z)$, as

\begin{equation}
\frac{\partial P_{\Phi} (k,z)} {\partial z}= \frac{\partial P(k,z)} {\partial z} + \frac{2}{(1+z)}.
\end{equation}

\begin{figure}
\includegraphics[width=6in, height=3in]{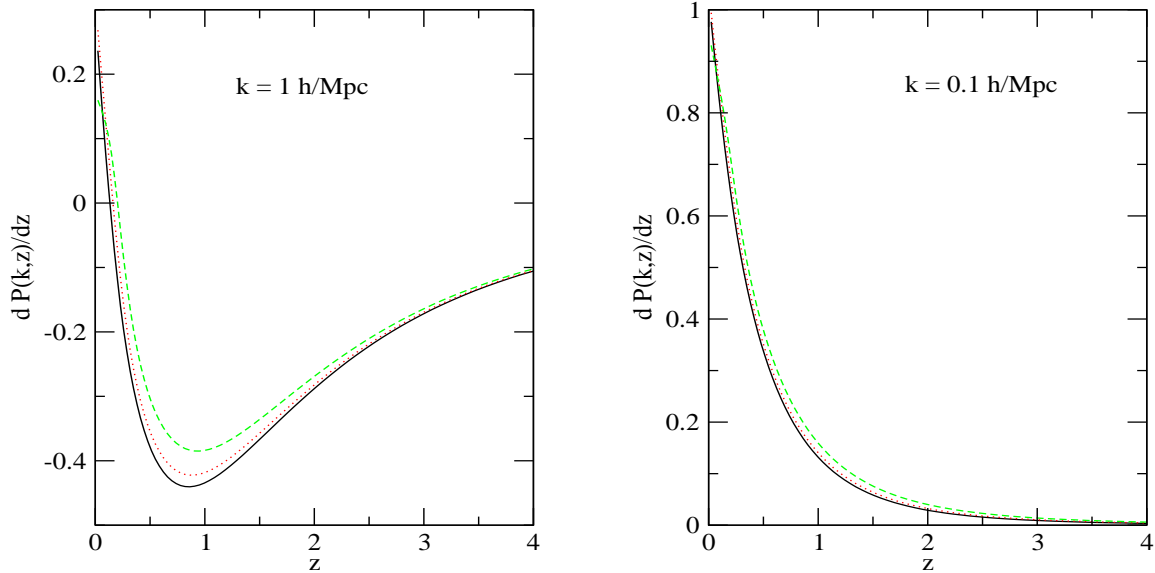}
\caption{\label{dPdz_model2}  Left panel: Evolution of nonlinear power spectrum for the model II at $k = 1$ h$Mpc^{-1}$, 
 where the solid (black), dot (red), and dashed (green) lines represent $\beta = $ 0.025, 0.05, and 0.1, respectively. 
 Right panel : Same that in the left panel, but at $k = 0.1$ h$Mpc^{-1}$.}
\end{figure}

As $P(k, z) \propto (1 + z)^{-2}$ for the linear matter power spectrum during the matter-dominated era,
$\partial P_{\Phi}(k, z)/ \partial z$ vanishes for this case, as expected. When the universe is dominated
by 'dark energy', the first term is still negative but becomes smaller than the second term, yielding
$\partial P_{\Phi}(k, z)/ \partial z > 0$. On the other hand, the Eq. (\ref{Pk_nonlinear})
shows that nonlinear evolution gives a term in $P(k, z)$ which goes as $(1 + z)^{-4}$,
and thus one obtains non-zero $\partial P_{\Phi}(k, z)/ \partial z$ even during the matter-dominated era.
The sign is opposite, $\partial P_{\Phi}(k, z)/ \partial z < 0$.
\\

\begin{figure}
 \includegraphics[width=6in, height=3in]{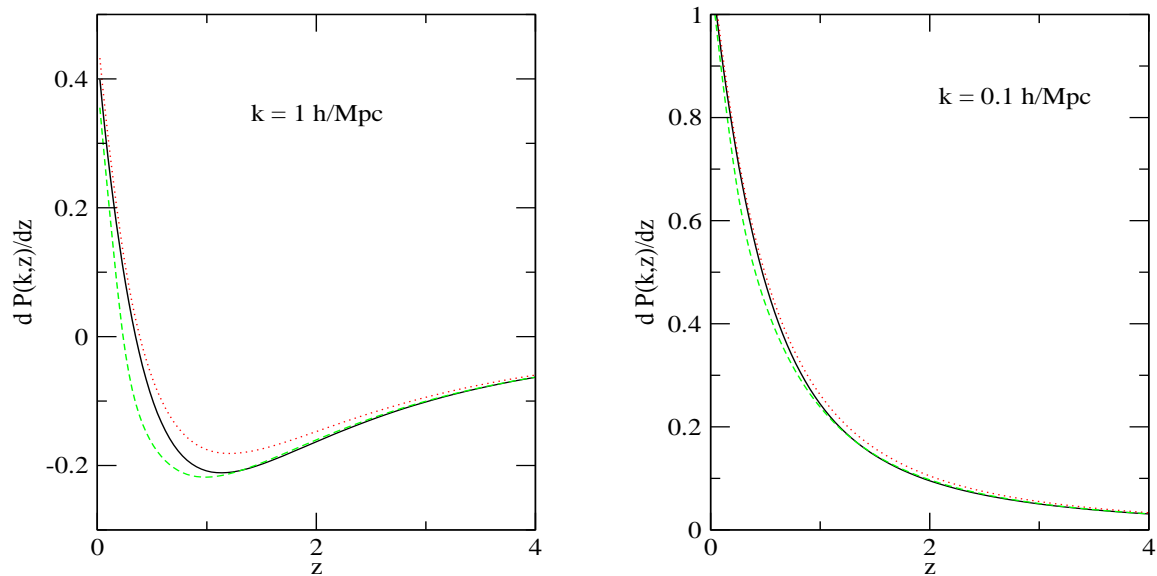}
 \caption{\label{dP_dz_model3} Left panel: Evolution of nonlinear power spectrum for the model III at $k = 1$ h$Mpc^{-1}$, 
 where the solid (black), dot (red), and dashed (green) lines represent $\beta = $ 0.025, 0.05, and 0.1, respectively. 
 Right panel : Same that in the left panel, but at $k = 0.1 \,$ h$Mpc^{-1}$. }
\end{figure}

Figure \ref{dPdz_model1} shows  $\partial P_{\Phi}(k, z)/ \partial z$ at $k = 1$ h$Mpc^{-1}$ (left panel) 
and $k = 0.1$ h$Mpc^{-1}$ (right panel) as a function of $z$ for the model I. The black, red, and green 
line represent $\beta = $ 0.025, 0.05, and 0.1, respectively.  
Note that the presence of matter creation at small scales decreases the effects nonlinears, where 
we see clearly the change in the sign of $\partial P_{\Phi}(k, z)/ \partial z$ at a moderate redshift 
when there is not or there is a small production of particles ($\beta < 0.05$). 
This regime is quite nonlinear (for $\beta < 0.05$).
Recall that the linear evolution due to 'dark energy' gives a positive contribution 
to $\partial P_{\Phi}(k, z)/ \partial z$, while the nonlinear 
evolution gives a negative contribution to $\partial P_{\Phi}(k, z)/ \partial z$.
Figure \ref{dPdz_model2} and \ref{dP_dz_model3} shows  $\partial P_{\Phi}(k, z)/ \partial z$ at $k = 1$ h$Mpc^{-1}$ 
and $k = 0.1$ h$Mpc^{-1}$ as a function of $z$ for the models II and III, respectively. Thus as in the 
figure \ref{dPdz_model1}, the black, red, and green lines represent $\beta = $ 0.025, 0.05, and 0.1, respectively.  
Note that the both models (II and III), the regime is quite nonlinear already at $k = 1$ h$Mpc^{-1}$, i.e, for this models
the rate of production particles not much affect the evolution of nonlinear power spectrum. Therefore, 
these models have a similar dynamic for the evolution of power spectrum when compared to the $\Lambda$CDM model.
\\

In this section, we can see that models where the dynamics associated with the production of 
particles is given by a $w_c = const$. e.g., model I, there is a great suppression  in the density 
contrast as we increase the rate production of matter. This becomes clear when analyzing the 
consequences in the power spectrum. On the other hand, a dynamic model for $w_c$, e.g, model II and III,
makes the evolution of perturbations of matter has a very reasonable behavior, i.e, similar to the model $\Lambda$CDM.

\section{Lensing-RS power spectrum}

Let us now illustrate the phenomenology described above for scenarios with matter creation, 
by computing explicitly the bispectrum signal as a function of the multipole $l$ described at detail in the section 2. 
On small multipole, the corresponding 
scales are outside the horizon, in linear regime, and $Q(l) > 0$. On the other hand, on sufficiently large multipoles, 
$Q(l)$ probes sub-horizon scales where the nonlinear regime dominates, and it has negative sign. The transition is 
located at the angular scale where the contributions from the scales in linear and nonlinear regime balance in the 
integrand, so that $Q(l)$ is zero.
\\

As mentioned previously, the essential ingredient of the lensing-RS bispectrum is the lensing-RS cross-power spectrum, 
$Q(l)$, defined by Eq. (\ref{Ql}). With  $\partial P_{\Phi}(k, z)/ \partial z$ computed, we are able 
to calculate the theoretical prediction of $Q(l)$ for models of matter creation. 
The sign of $\partial P_{\Phi}/ \partial z$ is determined by the balance of two competing contributions: 
the decaying of the gravitational potential fluctuations as $z \rightarrow 0$ and the amplification due to
nonlinear growth. Both of these are sensitive to the cosmological parameters, we thus expect $Q(l)$ is also.
The scale at which $Q(l)$ changes sign depends crucially on the scale at which the nonlinear growth overcomes 
the linear effect, making the the lensing-RS bispectrum sensitive to cosmological parameters governing the growth
of structure. We are interested in measuring the effects of creation of particles (characterized by $\beta$) on the 
lensing-RS bispectrum, so we fix some parameters. Fixed parameters for all analyzed cases: $H_0$ = 70 km/s/Mpc,
$\Omega_b=0.05$, $\Omega_{dm}=0.250$, $\Omega_{\Lambda}=0.70$, $\sigma_8 =0.834$, $n_s=0.96$ and $z_{ls}=1090$. 
\\

\begin{figure}
  \includegraphics[width=6in, height=3in]{Ql_model1.eps}
  \caption{\label{Ql_model1}  Left panel: Lensing-RS cross-power spectrum for the model I between $l \in [2,2500]$.
   Right panel: Same that in the left panel, but on the interval of $l \in [400,2000]$.
   The solid (black), dot (red), dashed (green), and dot-dashed (blue) lines represent $\Lambda$CDM ($\beta=0$), and $\beta$ = 0.025, 
   0.05 and 0.1, respectively.}
\end{figure}

\begin{figure}
 \includegraphics[width=6in, height=3in]{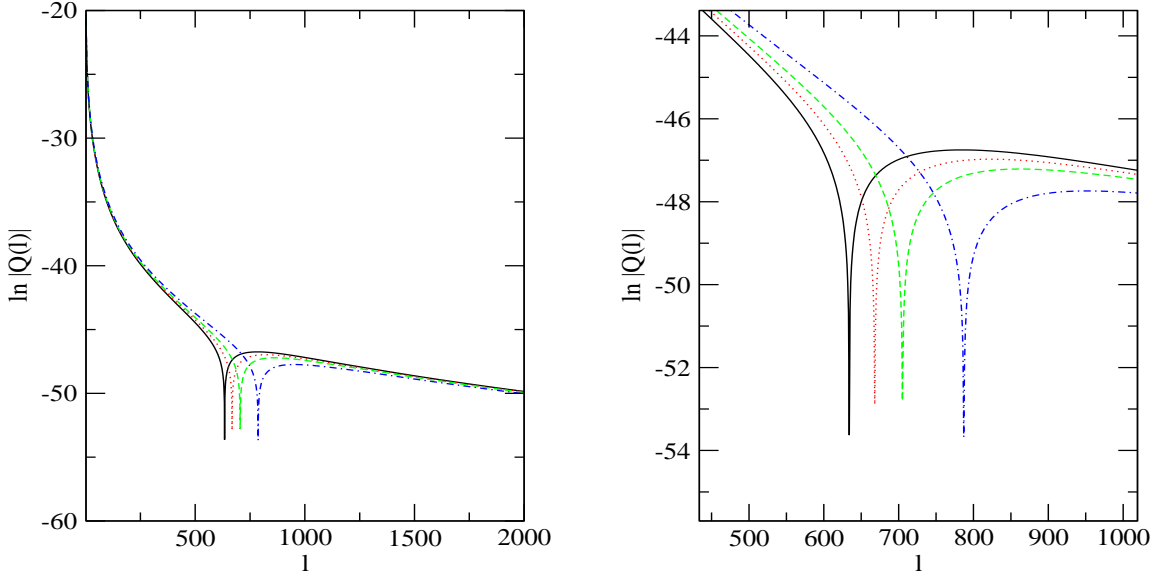}
 \caption{\label{Ql_model2} Left panel: Lensing-RS cross-power spectrum for the model II between $l \in [2,2000]$.
 Right panel: Same that in the left panel, but on the interval of $l \in [400,1000]$.
 The solid (black), dot (red), dashed (green), and dot-dashed (blue) lines represent $\Lambda$CDM ($\beta=0$), and  
 $\beta$ = 0.025, 0.05 and 0.1, respectively.}
\end{figure}

Figure \ref{Ql_model1} shows $|Q(l)|$ computed from Eq. (\ref{Ql}) for the model I. Where the black, red, green, and blue lines, 
represents $\Lambda$CDM ($\beta=0$), and  $\beta$ = 0.025, 0.05 and 0.1, respectively.
Note that the scale at which the nonlinear growth overcomes the linear effects occurs at smaller scales, 
when compared with $\Lambda$CDM. More specifically, the sign change in lensing-RS cross-power 
spectrum happens at $l \sim 640$ for $\Lambda$CDM, and at $l \simeq 1008$, $l \simeq 1270$, $l \simeq 1715$, 
for $\beta$ = 0.025, 0.05 and 0.1, respectively for the $\Lambda$CCDM model. 
\\

\begin{figure}
  \includegraphics[width=6in, height=3in]{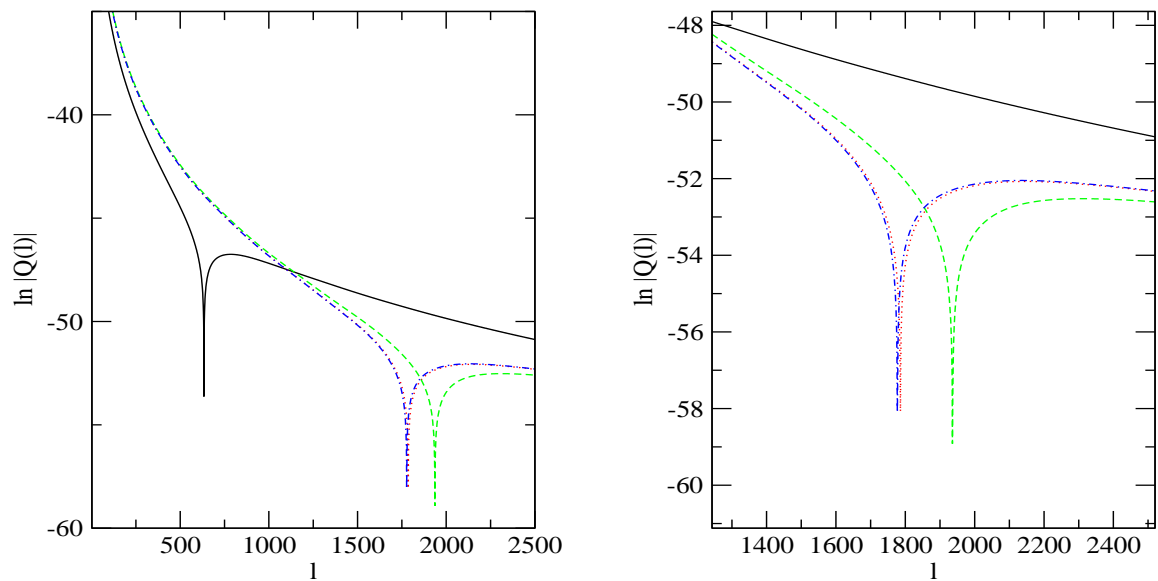}
  \caption{\label{Ql_model3} Left panel: Lensing-RS cross-power spectrum for the model III between $l \in [2,2500]$.
 Right panel: Same that in the left panel, but on the interval of $l \in [1300,2500]$.
 The solid (black), dot (red), dashed (green), and dot-dashed (blue) lines represent $\Lambda$CDM ($\beta=0$), and  $\beta$ = 
 0.025, 0.05 and 0.1, respectively.}
\end{figure}

Figure \ref{Ql_model2} and \ref{Ql_model3}, shows the effects for the models II and III, respectively. For the model II (III)
the sign change in lensing-RS cross-power spectrum happens at  $l \simeq 670 $ ($l \simeq 1780 $), $l \simeq 710 $  ($l \simeq 1940$), 
$l \simeq 790 $  ($l \simeq 1790 $), for $\beta$ = 0.025, 0.05 
and 0.1, respectively. Note that, thus as for the model I, for the model III  
the transition is located at very small scale angular. The model II, display small variations 
when compared as the $\Lambda$CDM model.
\\

The effects of the presence of matter creation is clear on lensing-RS bispectrum. 
Showing us that the scale where the nonlinear growth overcomes 
the linear effect depends strongly of particles creation rate $\Gamma$. 
We did not address here how much this effect is robust against variations in the other cosmological 
parameters, but the shift caused in the  lensing-RS bispectrum, due to the presence of matter creation is very clear.
In summary, the effects of the parameters $\beta$  on the lensing-RS bispectrum will go depend of dynamic associated with $\Gamma$.

\section{Conclusions}\label{sec:conclusions}

Understand the consequences of a particular physical mechanism, or cosmological model, on the process of structure formation is an essential ingredient in 
the search of a physical model that describes the observable universe.
Cosmological  models  driven  by  the  gravitational adiabatic  particle  production  have  been  intensively
investigated as a viable alternative to the $\Lambda$CDM cosmology \cite{Zeldovich,ccdm12}. Recently, Nunes and Pavon 
\cite{Dr1} explored the possibility that the EoS determined by recent observations \cite{ade,Scolnic} is in reality an effective EoS that
results from adding the negative EoS, $w_c$ (associated to the particle production pressure from the gravitational field acting on the vacuum) 
to the EoS of the vacuum itself, $w_{\Lambda}$. In the present paper, we investigate the
the implications of a continuous matter creation processes 
on structure formation in large and small scales, assuming three parametric models for the particle 
production rate given by Eq's. (\ref{proposta1}-\ref{proposta3}).
\\

We show as the matter creation affect the evolution of the perturbations of the matter the light of linear power spectrum of galaxy, 
and by a approximation in third order of perturbation to the nonlinear matter power spectrum. 
Note that when the dynamics associated with the particles prodution rate is given by a $w_c = const$, model I for example, 
there is a great suppression in the density contrast  as we increase the rate production of matter. 
When dynamic models are considered, models II and III, 
they are not observed large deviations in compared to theoretical prediction of $\Lambda$CDM model. We have shown 
via lensing-RS cross-power spectrum that the scale where the nonlinear growth overcomes 
the linear effect depends strongly of particles creation rate, this is, with the dynamic nature associated the $w_c$. 
\\

Obviously, phenomenological models of particle production different from the ones essayed here are also worth
exploring. More important, however, is to determine the rate $\Gamma$ using quantum field theory, but, as said above, 
this does not seem feasible until the nature of DM particles is found. Thus, hopefully through the development 
here performed within the context process of structure formation, help clarify the dynamic nature associated with 
the cosmological creation of particles.

\noindent 

\acknowledgments{ \noindent R.C.N. acknowledges financial support
from CAPES Scholarship Box 13222/13-9. The author is also very grateful to the anonymous referee for 
comments which improved the manuscript considerably.}
 
\end{document}